\documentclass[%
article,letterpaper,
superscriptaddress,
 amsmath,amssymb,
 aps,
]{revtex4}
\usepackage{tocvsec2}
\usepackage[utf8]{inputenc}
\usepackage{hyperref}
\hypersetup{
    colorlinks=true,
    linkcolor=blue,
    filecolor=magenta,      
    urlcolor=magenta,
    citecolor=blue
}
\usepackage[top=1in, bottom=1.25in, left=1.25in, right=1.25in]{geometry}
\usepackage{titlesec}
\usepackage{enumerate}
\usepackage{enumitem}
\usepackage{bm}
\usepackage{amsfonts}
\usepackage{amsmath}
\usepackage{amssymb}
\usepackage{etoolbox}
\patchcmd{\thebibliography}{\chapter*}{\section*}{}{}

\titlespacing*{\chapter}{0pt}{-50pt}{20pt}
\titleformat{\chapter}[display]{\normalfont\large\bfseries}{\chaptertitlename\ \thechapter}{20pt}{\Large}
\titleformat*{\section}{\fontsize{16}{20}\selectfont}

\def\ltap{\ \raise.3ex\hbox{$<$\kern-.75em\lower1ex\hbox{$\sim$}}\ }

\begin{document}

\title{Summary of the NuSTEC Workshop on Shallow- and Deep-Inelastic Scattering}


\author{C.~Andreopoulos} \affiliation{University of Liverpool, Department of Physics, Liverpool, United Kingdom}
\author{M.~Sajjad Athar} \affiliation{AMU Campus, Aligarh, Uttar Pradesh 202001, India}
\author{C.~Bronner} \affiliation{Kamioka Observatory, Institute for Cosmic Ray Research, University of Tokyo, Kamioka, Gifu, Japan}
\author{S.~Dytman} \affiliation{University of Pittsburgh, Pittsburgh, PA, 15260, USA}
\author{K.~Gallmeister} \affiliation{Institut f\"{u}r Theoretische Physik, Johann Wolfgang Goethe-Universit\"{a}t Frankfurt, Frankfurt a. M., Germany}
\author{H.~Haider} \affiliation{AMU Campus, Aligarh, Uttar Pradesh 202001, India}
\author{N.~Jachowicz} \affiliation{Department of Physics and Astronomy, Ghent University, B-9000 Gent, Belgium}
\author{M.~Kabirnezhad} \affiliation{University of Oxford, Oxford OX1 3RH, United Kingdom}
\author{T.~Katori} \affiliation{Queen Mary University of London, London E1 4NS, UK}
\author{S.~Kulagin} \affiliation{Institute for Nuclear Research, 117312 Moscow, Russia}
\author{A.~Kusina} \affiliation{Institute of Nuclear Physics PAN, Krakow, Poland}
\author{M.~Muether} \affiliation{Department of Mathematics, Statistics, and Physics, Wichita State University, Wichita, Kansas 67206, USA}
\author{S.~X.~Nakamura} \affiliation{University of Science and Technology of China, Hefei, 230026, China}
\author{E.~Paschos} \affiliation{Department of Physics, Technical University of Dortmund, D-44221 Dortmund, Germany}
\author{P.~Sala} \affiliation{CERN, CH-1211, Geneva, Switzerland}
\author{J.~Sobczyk} \affiliation{Wroclaw University, Faculty of Physics and Astronomy, Wroclaw, Poland}
\author{J.~Tena~Vidal} \affiliation{University of Liverpool, Department of Physics, Liverpool, United Kingdom}

\date{\today}

\begin{abstract}
The   \href{https://indico.cern.ch/event/727283/overview}{NuSTEC workshop} held at L'Aquila in October 2018 was devoted to neutrino nucleus scattering in the kinematic region where hadronic systems with invariant masses above the $\Delta(1232)$ resonance are produced: the so-called shallow- and deep-inelastic scattering regime. Not only is the physics in this kinematic region quite intriguing, it is also important for current and future oscillation experiments with accelerator and atmospheric neutrinos.
For the benefit of the community, links to the presentations are accompanied by annotations from the speakers.

\end{abstract}

\maketitle

\newpage

\tableofcontents

\newpage

\section{Foreword}

The neutrino-nucleus scattering regime starting just above the $\Delta (1232)$ resonance in invariant hadronic mass, $W$, covers two main kinematic regions labeled as Shallow Inelastic Scattering (SIS), with 1.4 GeV $< W <$ 2.0 GeV and Deep Inelastic Scattering (DIS), with $W > 2.0$ GeV (and $Q^2 > 1.0$~GeV$^2$).  Although these regions have been quite thoroughly investigated using \href{https://indico.cern.ch/event/727283/contributions/3102123/attachments/1731984/2799636/Keppel_nuSTEC.pdf}{e/$\mu$ scattering}  on nucleon ($N$) and nuclear ($A$) targets, they have hardly been covered either experimentally or theoretically in the case of  $\nu$-$N/A$ interactions. Aside from the inherent interest, present and future oscillation experiments have a significant number of events in these regions as illustrated by the sensitivity of NOvA (Sec.~\ref{sec:3A}) and atmospheric neutrino based experiments (Sec.~\ref{sec:3B}). The sensitivity of DUNE, expecting order 50\% of their events to have $W > M_{\Delta}$, is under study. 

SIS covers resonance excitation on the nucleon that, together with a non-resonant continuum, leads predominantly to single pion ($\pi N$) but also to $\gamma N$, $\eta N$ $K Y$, $\pi \pi N$, $\rho N$, ... final states. The currently available generators for $\nu-A$ event simulation use modified versions of the 
Rein-Sehgal model, first published in the early 80’s, (Sec.~\ref{sec:2A}, \ref{sec:2C}, \ref{sec:2D}, \ref{sec:2E}, \ref{sec:4B}) or, in the case of the Giessen BUU model (\ref{sec:2B}), rely on the MAID and SAID analyses of electron-nucleon and pion-nucleon scattering, respectively. The single pion production study of Hernandez, Nieves and Valverde, accounting for resonant and non-resonant mechanisms and their interference, has been modified to incorporate resonance excitation from the Rein-Sehgal model (Sec.~\ref{sec:4B}) and extended to higher invariant masses within the Regge approach  (Sec.~\ref{sec:5C}). The dynamical couple channel approach developed in Osaka, Sec.~\ref{sec:4A}, has predictions not only for $\pi N$ but for other meson-baryon final states.  The procedure used by simulation programs to model the non-resonant pion contribution to SIS relies on the  \href{https://indico.cern.ch/event/727283/contributions/3102125/attachments/1732202/2800115/BY_Nuscatt-18.pdf}{Bodek-Yang} model that simulates the inclusive non-resonant contribution without giving information on particular pion charge states.  

Relating the SIS and the DIS regions is the goal of quark-hadron duality that connects the DIS degrees of freedom (quarks and gluons)  with the SIS ones (hadrons).  Again, duality has been extensively studied both experimentally and theoretically in \href{https://indico.cern.ch/event/727283/contributions/3102166/attachments/1732853/2801414/duality_nustec_christy.pdf}{e/$\mu$–N/A}  scattering.  However, with the current singular lack of $\nu-N/A$ experimental data covering this region, the study of duality in neutrino scattering has been limited to purely model-dependent studies (Sec.~\ref{sec:5A}).  Using the above summarized models for resonant and non-resonant pion production there are variations in the results of duality tests ranging from obvious problems to conditional acceptance.  Since duality is an accepted concept in the strong and electromagnetic interaction landscape, the problems with tests of duality with neutrinos can be due to faulty resonance models and/or incorrect non-resonant pion modeling or, perhaps, the application of duality tests with neutrinos could be different than for electromagnetic scattering.  The bridge from SIS to DIS region is also an excellent kinematic region to probe non-perturbative QCD phenomena highlighted in presentations \ref{sec:5B} and \ref{sec:5C}.

In the DIS region there have been recent extractions of nuclear parton distribution functions (nPDFs) that suggest the existing high-statistics $\nu-A$ experimental results lead to quite different nPDFs than the many experimental results from e-A scattering (Sec.~\ref{sec:6C}). This is not surprising for the low-$x_{Bj}$ region since there has long been theoretical speculation that the inclusion of the axial-vector current does indeed change shadowing for $\nu-A$ scattering.  In directly calculating the nuclear structure functions, rather than extracting the nPDFs, these derivations at low-$x_{Bj}$ are also detected although at a somewhat smaller scale (Sec.~\ref{sec:6A} and \ref{sec:6B}).  This, in turn, suggests that further thought is needed to preserve the concept of {\it universal} nuclear parton distribution functions.

Another important subject is the neutrino hadronization problem. In fact, hadronization has been first carefully studied by neutrino interactions in 80's bubble chamber experiments. However, the simulation of hadronization in generators is not trivial (Sec.~\ref{sec:7A}, \ref{sec:7B}, \ref{sec:7C}). This subject has been neglected for many years due to lack of interest. Current generators do not support the systematic error estimation necessary for experiments above the few-GeV region (\ref{sec:7D}). These errors could be important for future oscillation experiments dominated by systematics. 

\vspace{0.25cm}
\hspace{3cm} Summary Editors: Luis Alvarez Ruso, Teppei Katori and Jorge G. Morf\'{i}n 

\newpage

\section{Generator / Transport Treatments of the SIS and DIS Region}
\subsection[Status of Resonances in GENIE. Emphasis on Resonances above the $P_{33}(1232)$ – \\Steve Dytman]{Status of Resonances in GENIE. Emphasis on Resonances above the $P_{33}(1232)$ – \\Steve Dytman \label{sec:2A}
[\href{https://indico.cern.ch/event/727283/contributions/3102124/attachments/1731994/2799654/reinsehgal-dytman-oct18.pdf}{Presentation}]}

GENIE uses a mixed treatment for the SIS kinematical region.  Ingredients include a resonance model (either Rein-Sehgal or Berger-Sehgal) which has many resonances and a full model by Bodek and Yang.  The resonant models apply to $W$ up to about 2.3 GeV; the Bodek-Yang model was designed to average over resonances and can be used down to $\pi N$ threshold.  In v2 and earlier, there is a cutoff of $W=1.7$ GeV.  Above that cutoff, the Bodek-Yang model is used for the full response.  Below that cutoff, the Bodek-Yang result is scaled to match the deuterium pion production data when added to the resonant cross section.  The resonance masses and widths are regularly updated to match PDG recommendations; this method produces good agreement with data.

Although this is reasonable given the inputs, it is not fully satisfactory.  The Bodek-Yang model includes both resonant and nonresonant contributions, so in some sense the resonance model is over-counted and the nonresonant model is under-counted.  Proper interference between resonant and nonresonant processes is not possible at present.  Various improvements in models are coming together, e.g. getting the vector form factors from electron scattering data.

The SIS region is very interesting.  It can be described using resonance language or DIS language.  The $\Delta$(1232) region is very well understood in a model using hadronic degrees of freedom.  Theoretically, the physics is well understood by including 1-pion Born diagrams interfering with a resonance diagram.  The 2-pion Born diagrams are much more complicated and may be difficult to implement properly.  The DCC model (see Satoshi Nakamura's \href{https://indico.cern.ch/event/727283/contributions/3160945/attachments/1732798/2801299/s.nakamura_nustec.pdf}{presentation}) has a model for the 2-pion total response using the resonance picture.  DIS models include all processes but are mostly applicable to inclusive data and have trouble getting the correct kinematic distributions of hadrons in the final state.  Future efforts should aim at getting a proper mix of these two different pictures.



\subsection[GiBUU Treatment of DIS – Kai Gallmeister]{ GiBUU Treatment of DIS– Kai Gallmeister \label{sec:2B}
[\href{https://indico.cern.ch/event/727283/contributions/3102157/attachments/1731860/2799423/Talk1.pdf}{Presentation}]}

GiBUU uses a multi-resonance model for low energetic electron-nucleon, photon-nucleon or neutrino-nucleon interactions. 
The vector couplings of resonances and background
contributions are all fixed from electron-nucleon data via the MAID analysis. The axial couplings are fixed by PCAC.
Cross sections on the nucleus are then obtained by Lorentz-boosting the elementary
cross sections on the nucleon into the rest-frame of the Fermi-moving nucleons in the nucleus.

Higher energetic interactions are described via the PYTHIA framework (which is a part of GiBUU). Here both the cross section calculation and also the event generation are done by this external package. 

Relying on version PYTHIA v6.2, electron and photon induced interaction are done using the sophisticated implemented prescription. 
For neutrino induced interactions, the basic fermion+fermion $\to$ fermion+fermion
prescription is used to calculate the
inclusive cross section; the fragmentation into the final state hadrons is achieved by
the string fragmentation mechanisms built into PYTHIA. A careful choice of form factors/suppression factors yields
very good results for (e,A) and ($\nu$,A) reactions.


\subsection[Tuning the Pion Production with GENIE version 3  – Julia Tena Vidal]{Tuning the Pion Production with GENIE version 3  – Julia Tena Vidal \label{sec:2C}
[\href{https://indico.cern.ch/event/727283/contributions/3102158/attachments/1731996/2799656/PionProductiontune_Geniev3_JTena_NuSTEC.pdf}{Presentation}]}

This talk focuses on the description of shallow inelastic scattering (SIS) region in GENIE
\cite{GENIE}
and describes a new tune on pion production at the free nucleon level using deuterium data from the ANL-12ft, BNL-7ft, BEBC and FNAL 15-ft bubble chamber experiments
\cite{ANL,BNL,BEBC,FNAL}.

The SIS region is fundamental to describe single and double pion production mechanisms and yet there is not a single model that describes simultaneously resonant and non-resonant interactions. The modelling of this transition region is left to the generators and a number empirical models are used to achieve this goal. In GENIE, there are two models available to describe RES interactions, either Rein-Sehgal or Berger-Sehgal~\cite{RS,BS}. From the $\Delta$ peak to the pure DIS regime, $W>2$ GeV, a non-resonant background needs to be added,
  \begin{equation}
     \frac{d^2\sigma^{SIS}}{dQ^2dW}=\frac{d^2\sigma^{RES}}{dQ^2dW}+\frac{d^2\sigma^{Non-RES}}{dQ^2dW}\,.
 \end{equation}
 
Particularly in GENIE, the non-resonant background is the extrapolation of the DIS cross section at lower $W$ scaled by some multiplicity functions, $f_m=R_m\cdot P^{had}_m$.  The end of the transition region is determined by $W_{cut}$, with $W>W_{cut}$ being the pure DIS regime,

 \begin{equation}
     \frac{d^2\sigma^{RES}}{dQ^2dW}=\sum_k\left(\frac{d^2\tilde{\sigma}^{RES}}{dQ^2dW}\right)_k\cdot\Theta(W_{cut}-W)
 \end{equation}
 
  \begin{equation}
     \frac{d^2\sigma^{Non-RES}}{dQ^2dW}=\frac{d^2\tilde{\sigma}^{DIS}}{dQ^2dW}\cdot\Theta(W_{cut}-W)\cdot\sum_mf_m
 \end{equation}

$\tilde{\sigma}^{RES}$ is the cross section for RES given a specific model, either Rein-Sehgal or Berger-Sehgal. For this specific tune, the Berger-Sehgal model has been implemented. DIS is modelled with the Bodek-Yang model \cite{BodekYang} that is described in more detail in the associated \href{https://indico.cern.ch/event/727283/contributions/3102125/attachments/1732202/2800115/BY_Nuscatt-18.pdf}{presentation}. $R_m$ are tunable ad hoc parameters that depend on the neutrino flavour, on the multiplicity, $m$, of the final state to be $m+2,\,3$ and on the ID of the initial state nucleon. Moreover, $P^{had}_m$ is the probability of the final state to be $m+2,\,3$ coming from the hadronization model. 

GENIE has addressed the modelling of pion production at the free nucleon level with a new tune using deuterium data from ANL-12ft, BNL-7ft, BEBC and FNAL-15ft bubble chamber experiments. The shallow inelastic scattering region has been tuned against $\nu_\mu$ and $\bar{\nu}_\mu$ CC inclusive, quasi-elastic, one pion and two pion integrated cross sections as a function of $E_\nu$. Not all the available historical data has been used for the fit, as some of the datasets have been reanalyzed \cite{WCR}. Particularly for ANL-12ft and BNL-7ft, reanalyzed data has been used when available. Quasi-elastic data has been introduced to the fit to better constrain the flux of each experiment. 

In order to tune the SIS region against free nucleon data, a total of nine parameters are included in the fit: $M_A^{RES}$, $M_A^{QE}$, R-vp-CC-m2, R-vn-CC-m2, R-vp-CC-m3, R-vn-CC-m3, DIS-XSecScale, RES-XSecScale and $W_{cut}$. Previous analysis to extract $M_A^{RES}$ and $M_A^{QE}$ from $\nu_\mu$ CC bubble chamber data as a function of $Q^2$ and $W$ have been used as priors during the fit. Due to correlations between datasets coming from the same experiments, an extra nuisance parameter per experiment has been considered in the fit. 

The global fit describes both inclusive and exclusive cross sections simultaneously, improving the agreement for $\nu_\mu$CC p$\pi^+$, n$\pi^+$, p$\pi^0$ and p$\pi^+\pi^-$ cross sections on free nucleon. Tensions between inclusive and exclusive data have been re-encountered, showing a decrease in the inclusive cross section at the 1-10 GeV energy region when fitting exclusive data. This effect has been observed for one pion production channels, and in particular, the non-resonant background contribution is now lower. The prediction for two pion production mechanisms is in better agreement with data after the global fit for the $\nu_\mu n \rightarrow \mu^- p \pi^+ \pi^-$ channel. 

\subsection[DIS Event Generation in NEUT - Christophe Bronner]{DIS Event Generation in NEUT - Christophe Bronner \label{sec:2D} 
[\href{https://indico.cern.ch/event/727283/contributions/3102160/attachments/1732047/2799754/NEUT_DIS.pdf}{Presentation}]}

In this presentation, I described the way deep-inelastic (DIS) events are simulated in the NEUT neutrino interaction generator\cite{NEUT}, and recent updates related to the simulation of those events. This presentation was based on version 5.4.0 of NEUT.\\

Generators use a superposition of different models to simulate events, and slide 2 presents the models used in NEUT to generate events above the pion production threshold. There are 2 different regimes depending on the value of the hadronic invariant mass W:
\begin{itemize}
\item below 2 GeV/c$^{2}$, events with only one particle produced on top of the outgoing lepton and baryon are simulated using resonant models, while the events with 2 or more additional particles are simulated using a custom DIS model.
\item above 2 GeV/c$^{2}$, all the events are simulated using a DIS model, based on the external generator PYTHIA\cite{PYTHIA} v5.72  
\end{itemize}
There are therefore two different DIS models used in NEUT, which we will refer to as the low and high W modes.

Slide 3 presents the different steps to generate a DIS event in NEUT. The calculation of the cross-section, which tells us how many events should be generated is done first, then actual events are generated. In the case of the low W model, variables describing the interaction at a more global level ((x,y) or (W,Q$^{2}$)) are generated first, then the hadronic system is simulated based on the value of W generated. In the case of the high W model the 2 steps are done simultaneously by PYTHIA.\\

Slides 4 and 5 introduce the way the double differential cross-section in x and y is calculated. This calculation is an important step in the simulation, as this double differential cross-section is what allows both to compute the total cross-section, and to generate the global variables in the low W mode. The main sources of uncertainty at this level are the relations between the different structure functions used to relate $F_{2}$, $F_{4}$ and $F_{5}$ to $F_{1}$, as well as the Parton Distribution Functions (PDFs), which are the inputs used in this calculation. We use the modified GRV98\cite{GRV98} PDFs from the Bodek-Yang model\cite{BYGRV98}.\\

Slides 6 to 13 focus on the low W DIS mode. This mode was significantly updated since NEUT 5.3.2, after it was noticed at NUINT2015\cite{NUINT2015} that although the low W DIS modes of NEUT and GENIE\cite{GENIE} followed similar approaches, noticeable differences could be seen between the W and Q$^{2}$ distributions of the events they produced. After those updates, the predictions from the 2 generators are in good agreement when generating events in the same regions of phase space.

To avoid double counting between DIS and resonant events in this low W region, only DIS events with 2 or more hadrons on top of the the outgoing lepton and baryon are kept. This gives a particular importance to the multiplicity model, which gives the probability to produce a given number of hadrons as a function of W and the incoming neutrino and target nucleon types. Three such models are implemented in NEUT 5.4.0, and the comparison between the events generated in each case allows us to see that the choice of the multiplicity model has a significant impact on the total cross-section, the W distribution and the kinematics of the hadrons produced for the low W DIS events.\\

Slide 14 and 15 briefly describes the high W mode. In this case PYTHIA 5 is used with only a few settings changed to produce the events. The types and 4-momenta of the incoming neutrino and target nucleon are provided as inputs to PYTHIA, and events are generated until an event of the right type (charged or neutral current) and  W$>$2 GeV/c$^{2}$ is obtained. Other generators use the more recent PYTHIA 6, and tests were done to evaluate the impact of using an older version of PYTHIA in NEUT. It was found that the change from version 5 to version 6 did not have any impact on hadronization. The main difference concerns the generation of the global variables (W,Q$^{2}$)): PYTHIA 5 can be used to generate those variables, but not PYTHIA 6. Generators using PYTHIA 6 therefore need to first generate W and Q$^{2}$ based on the double differential cross-section in x and y, and pass the value of W obtained as an input to PYTHIA.\\

Slide 16 describes future updates: the recent updates on the DIS modes were focusing on charged-current events, work is now on-going to improve the simulation of neutral current events.

\subsection[NuWRO SIS/DIS Model – Jan Sobczyk]{NuWRO SIS/DIS Model – Jan Sobczyk \label{sec:2E}
[\href{https://indico.cern.ch/event/727283/contributions/3102159/attachments/1731863/2799427/nuint18-nuwro-sis-dis.pdf}{Presentation}]}

The presentation starts with a general information about NuWro (slides 2-4). 

Slide 5 is a description of NuWro RES/DIS transition region (technically it is contained in RES covering $W\leq 1.6$~GeV). The transition region is $W\in (1.3, 1.6)$~GeV. NuWro RES model is a combination (a incoherent sum) of $\Delta$ excitation and nonresonant background. $\Delta$ excitation model parameters are taken as a fit to ANL/BNL data in $p\ \pi^+$ channel. Non-resonant background is technically modelled as a fraction of DIS with weights - functions of $W$ adjusted to the data  in the remaining channels.

Slides 6-11 are devoted to DIS, mostly to hadronization model. Slide 10 shows specific choices of PYTHIA parameters (the goal was to get a good agreement with the charged multiplicity data) and slide 11 results for the charged hadron multiplicity, taken from the paper written by paper of Kuzmin and Naumov. 

Slide 12  is a  resume of NuWro RES and DIS models.

Remaining slides contain a selection of figures showing NuWro (version 18.02 is used) performance compared to the data. 

Slides 14-15 show MINERvA inclusive cross section data. The data is both for neutrinos and antineutrinos, and also for the cross section ratio. In all the cases they are given as functions of (anti)neutrino energy.

Slides 16-19 show T2K inclusive flux integrated cross sections in muon bins (angle and momentum).

Slides 20-21 show MINERvA "low recoil" results in bins of reconstructed three-momentum and "available energy".


\subsection[Generator Comparisons: SIS/DIS region - Christophe Bronner]{Generator Comparisons: SIS/DIS region - Christophe Bronner \label{sec:2F} 
[\href{https://indico.cern.ch/event/727283/contributions/3102161/attachments/1732050/2799758/Generators.pdf}{Presentation}]}

In this presentation, we look at the differences in how the three main generators used by neutrino oscillation experiments (NEUT~\cite{NEUT}, GENIE~\cite{GENIE} and NuWro~\cite{NuWro}) treat the SIS/DIS region, and compare their predictions in this region for a number of kinematical variables and particle multiplicities. We first try to present the sources of the differences between those predictions, and then show comparisons for interactions of neutrinos of different energies on different nuclear target.

The SIS/DIS region is assumed here to be defined by an hadronic invariant mass W larger than 1.7 GeV/c$^{2}$, and the study was limited to charged-current interactions of muon neutrinos and antineutrinos. In practice, NEUT 5.4.0, GENIE 2.12.10 and NuWro 18.02.1 were ran with their default settings, and only charged current events of the DIS and resonant modes with W$>$1.7 GeV/c$^{2}$ were kept.\\

Above the pion production threshold, the generators use different models depending on the value of W. The general pattern is to use resonant models at low W (possibly with a non resonant background described by a custom DIS model) and to use a purely DIS model at high W, based on the PYTHIA\cite{PYTHIA, PYTHIA6} generator. However, the number of resonances considered, the number of intermediate steps in this transition, as well as the values of W at which the transitions occur differ between the generators. Those differences are clearly visible when looking at the W distributions of events produced by the 3 generators for the interactions of neutrinos of a few GeV.\\

Differences can be seen between the generators even in regions of W where they use similar models. We can first look at the region 1.7 GeV/c$^{2}$ $<$ W$<$2.0 GeV/c$^{2}$, where both NEUT and GENIE use a custom DIS model (referred to as `low W' models in this presentation). The two generators use a similar approach and the same inputs to generate the Bjorken x and y variables describing the event (those variables are equivalent to W and Q$^{2}$, and determine the outgoing lepton kinematics as well as the energy available to build the hadronic system). However, even if we look at interactions on free nucleons to avoid differences due to nuclear models and final state interactions (FSI), the predictions of the 2 generators used with their default settings differ for those variables. It is shown that those differences come from 3 (rather technical) differences: the scaling variable used for relations between the structure functions, the model used for the hadronic multiplicities, and the fact that GENIE computes separately the structure functions for interactions on each type of quark while NEUT does not.  

On slides 11 to 13, we look at the differences in the hadron kinematics, and more particularly the leading pion kinetic energy $T_{\pi}$, between the GENIE and NEUT low W models. Those differences are found to come from 3 sources: differences in the hadron multiplicity models, the use in GENIE of experimental data to simulate the kinematics of the outgoing nucleon, and differences in the FSI model. When the same hadron multiplicity model is used in GENIE and NEUT, the difference in  $T_{\pi}$ predictions come mainly from the differences in FSI models.\\

When comparing the predictions for x and y in the high W region (W $>$3.0 GeV/c$^{2}$, where all the generators use a DIS model based on PYTHIA), good agreement is seen between GENIE and NuWro, while NEUT predictions look different. The reason is believed to be that GENIE and NuWro generate (x,y) based on the double differential cross-section in x and y, while NEUT uses PYTHIA 5 for this. We can note that the differences in parton distribution functions used by GENIE and NuWro (from the Bodek-Yang models based on GRV94 and GRV98 respectively\cite{BYGRV94, BYGRV98}) do not seem to have an impact on those predictions.\\

The rest of the presentation is made of comparisons of the predictions of the generators for different variables (W, Q$^{2}$, lepton angle and momentum and multiplicities of charged hadrons, pions, and neutral pions) for the interactions of neutrinos of different fixed energies on different nuclear targets. All the effects described previously are present simultaneously so that it is not easy to interpret the differences, but some surprising features are seen in the predictions of the lepton angle and momentum (slide 19 and 20). 

\newpage

\section{Sensitivity of Oscillation Parameters to the SIS and DIS Region}
\subsection[Deep Inelastic Scattering Impact on NOvA – Mathew Muether ]{Deep Inelastic Scattering Impact on NOvA – Mathew Muether \label{sec:3A}
[\href{https://indico.cern.ch/event/727283/contributions/3102148/attachments/1731862/2800125/NuSTECDIDOct2018_Final2.pdf}{Presentation}]}

This presentation describes the impact of shallow (SIS) and deep (DIS) inelastic neutrino scattering on measurements from the NOvA Neutrino Experiment and future efforts from NOvA to measure interactions in this region. NOvA is a long-baseline neutrino oscillation experiment which uses the Fermilab NuMI beam to expose a pair (near and far) of active tracking calorimeters to muon-neutrinos and measure electron-neutrino appearance and muon-neutrino disappearance. The neutrino spectrum at NOvA peaks at 2 GeV with the bulk of the exposure ranging from 1-3 GeV with a non-negligible higher energy tail. 

NOvA uses the large sample of neutrino interactions in the near detector to tune a neutrino interaction model using GENIE for use in oscillation and additional interaction measurements. NOvAs find an optimal tune by making the following changes to the base GENIE model: the Valencia RPA model of nuclear charge screening is applied to quasi-elastic and resonance processes \cite{Muether2}, 10 percent increase in non-resonant DIS at high W, reduced  normalization of GENIE non-resonant single pion production with W $<$ 1.7 GeV by 57 percent \cite{Muether4} (this is applied to neutrinos but not antineutrinos), add tuned empirical MEC events \cite{Katori:2013eoa}. With these modifications the NOvA near detector data matches well with the model. Some disagreement is visible but it is notable that the DIS and RES contributions strongly overlap making their independent contributions to the disagreement difficult to determine. 

Systematics which are included this fits are described. GENIE includes DIS normalization systematics of 50\% for 1 and 2 pion final states in events with W $<$ 1.7 GeV. NOvA expands these to apply to final states with any number of pions. We also increased the range of the systematic to apply up to a W of 3 GeV as the discontinuity of 50\% $<$ 1.7 GeV and 0\% $>$ 1.7 GeV seems unphysical, even though we know higher energy regions are better constrained. We feel in general untrusting of this region in GENIE, hence the large uncertainties, and would greatly appreciate a closer look from the community at the model and the systematics GENIE provides.

NOvA has observed a data/MC discrepancy in the low track- length, high y region of numu- selected ND events. CVN \cite{Muether1} (NOvA's event classifier) recovers many of the low track-length events but due to this discrepancy we continue to apply a muon selection using Remid. Resolving this discrepancy and relaxing ReMID requirements would boost available analysis statistics.

NOvA is currently working on a set of cross-section measurements which may help better understand DIS. 

\subsection[SIS/DIS Interactions and Uncertainties in Atmospheric Oscillation Analysis – \\Christophe Bronner]{SIS/DIS Interactions and Uncertainties in Atmospheric Oscillation Analysis – \\Christophe Bronner \label{sec:3B} 
[\href{https://indico.cern.ch/event/727283/contributions/3102149/attachments/1732052/2799761/Atmo.pdf}{Presentation}]}

This presentation looks at the importance of the modelization of SIS/DIS interactions for the study of the oscillations of atmospheric neutrinos, through the particular example of the search for the neutrino mass hierarchy in the Super-Kamiokande (SK) experiment. Other experiments studying or planning to study the oscillations of atmospheric neutrinos (IceCube, KM3NET) use different analysis and reconstruction methods, but some of the problems presented here will be relevant for them as well.\\

A neutrino produced in a given flavor can after some propagation be detected as a neutrino of a different flavor. This oscillation phenomenon is possible since the neutrino mass eigenstates, which are the propagation eigenstates, are different from the flavor eigenstates in which the neutrinos interact through charged current weak interactions. One of the three main open questions in the study of neutrino oscillations is the neutrino mass hierarchy (MH): is the third mass eigenstate heavier (normal hierarchy - NH) or lighter (inverted hierarchy - IH) than the two other mass eigenstates?

Atmospheric neutrinos can be used to determine the MH by taking advantage of the modifications in the oscillation pattern coming from the propagation through matter: oscillations of neutrino propagating through matter differ from oscillations in vacuum, and those differences depend on the MH. In particular, a resonance is expected to occur in the oscillation from the electron to the muon flavor, for neutrinos if the hierarchy is normal and for antineutrinos if it is inverted. Based on the current knowledge of oscillation parameters, the resonance should happen for upward going neutrinos or anti-neutrinos of energies 2-10 GeV. At those energies a significant fraction of the neutrino interactions are of the SIS/DIS type.\\

Super-Kamiokande (and the future Hyper-Kamiokande) are Water Cerenkov detectors. This type of detector is very good at separating charged current interactions of muon neutrinos from those of electron neutrinos, but cannot distinguish on an event by event basis the interactions of neutrinos from those of anti-neutrinos. This leaves two handles to try to determine the MH using atmospheric neutrinos:
\begin{itemize}
\item the size of the resonance: since both the flux and the cross-sections are larger for neutrinos than for antineutrinos, we expect a larger signal if the hierarchy is normal.
\item using the differences between the interactions of neutrinos and anti-neutrinos to separate them on a statistical basis, and build samples enriched in either $\nu_{e}$ or $\bar{\nu}_{e}$ events.
\end{itemize}

It is difficult in practice to only use the size of the resonance to determine the MH: the flux and cross-sections are subject to non-negligible systematic uncertainties, and the expected size of the resonance signal also depends on the value of the other oscillation parameters, in particular $\sin^{2}(\theta_{23})$ which is not well constrained. For this reason, we use in the SK oscillation analysis a statistical separation based on the fact that the DIS interactions of anti-neutrinos have on average larger Bjorken y than the ones from neutrinos. We therefore expect that for the interactions of neutrinos the transverse momentum will be larger, and more hadrons will be produced, visible in the detector either as ``rings'' or as Michel electrons coming from the decays of charged pions. \\

The remainder of the presentation describes the DIS related uncertainties which affect the analysis, how they are treated in the current SK analysis and areas where their modelization could be improved. The two handles listed above are predominantly affected by different uncertainties. For the size of the resonance, it is mainly the uncertainty on the total cross-section that will affect the number of events expected in this region. For the statistical separation of neutrinos and anti-neutrinos, details of the modelization of the DIS interactions, in particular the double differential cross-section in (x,y) or (W,Q$^{2}$) and the properties of the hadronic system will be more important. Historically, systematic uncertainties in the SK analysis have focused on the cross-section part, and uncertainties on the topology of the events have only recently begun to be studied in more details.\\

For DIS interactions, the cross-section as a function of energy is obtained by integrating the double differential cross-section in x and y over those two variables. $d^{2}\sigma/dxdy$ is parameterized in terms of structure functions, and the main sources of uncertainties on the cross-section will be the parton distribution functions (PDF) used to compute the structure functions, and the relations between those structure functions, as in practice only F$_{2}$ and xF$_{3}$ are computed and F$_{1}$, F$_{5}$ are deduced from F$_{2}$ while F$_{4}$ is taken to be zero.

The PDFs can be computed in QCD, with free parameters determined by a fit to the data. This can only be done however for large enough Q$^{2}$, with a threshold typically around 1 GeV. This limitation is problematic for neutrino oscillation experiments: a significant fraction of the DIS interactions they observe have a Q$^{2}<$1 GeV. Bodek and Yang have devised a set of corrections for the GRV98 LO PDFs that allow to evaluate them at low Q$^{2}$ (see ``Current status of the Bodek-Yang model'' in this workshop). For this reason, neutrino interaction generators still use the GRV98 LO PDFs~\cite{GRV98}, although more recent PDFs are available. The Bodek-Yang model contains free parameters, which are determined by a fit to electron scattering and photo-production data.

For the uncertainties on the Bodek-Yang models, two types of approaches have been used in neutrino oscillation analyses: using errors on the values of the model parameters, or simply considering that the difference between the GRV98 LO PDFs and their modified version in the Bodek-Yang model represents the uncertainty on the PDFs as a function of Q$^{2}$. In the case of the SK analysis, we found it difficult to use uncertainties on the parameters values with the information currently available for the version of the model used in the analysis~\cite{BodekYang}, and decided to use the difference between the PDFs in the two cases as our uncertainty. The goal for the future is to use uncertainties on the parameter values instead.

The analysis includes two other systematic parameters affecting directly the DIS cross-section. The first one is referred to as ``DIS model uncertainty'', and corresponds to the difference in the predicted cross-section between our nominal model and an alternative model based on conventional Regge theory, the CKMT model~\cite{CKMT}. As this alternative model is not really used anymore, we are considering changing for a comparison between the predictions of the Bodek-Yang model and those obtained using more modern PDFs for the  Q$^{2}>$1 GeV region. The second uncertainty is a normalization uncertainty, coming from the difference between the NEUT CC inclusive cross-section prediction and the world average measurement (from PDG17\cite{PDG2017}) on the range 30-200 GeV.\\

A serious difficulty faced by experiments studying neutrinos in the range 2 to 10 GeV is that it corresponds to the transition between resonant and DIS interactions. Neutrino interaction generators use a set of different models to simulate interactions, using resonant models at low W and exclusively DIS models for large enough W, but there are no clear prescriptions on how to make the transition from resonant models to DIS models in the intermediate region. In the case of the SK analysis, resonant and DIS interactions are expected to look different in the detector, so the uncertainty on the transition from one regime to the other could create a systematic uncertainty on the statistical separation between neutrinos and anti-neutrinos in the resonance region.

In practice, the generators end up using a superposition of the two types of models in this region, relying on criteria based on the number of particles produced in the events to determine which model to use and to avoid double counting of the cross-section. For example, NEUT will use in the region W$<$2 GeV a DIS model to simulate events where at least 2 hadrons on top of the outgoing baryon are produced, and exclusive resonant models to simulate the other events. This ends up adding uncertainty on the cross-section in this region: the cross-section for the resonant modes is computed directly based on those models, while the cross-section for the low W DIS mode used here is obtained from the total DIS cross-section multiplied by a factor corresponding to the probability to have two or more hadrons produced at this energy. There are significant uncertainties on the hadron multiplicity model used to evaluate this probability, and this translates into an uncertainty on the total cross-section in this region. Additionally, this hadron multiplicity model is a key part of the custom low W DIS models used in NEUT and GENIE, and  affects other aspects of the simulation of DIS events in the transition region, such as the W distribution and the hadron kinematics. It would therefore be important to improve those models, but it is difficult in practice as they are based on fits of hydrogen and deuterium bubble chamber data, which seem to give different results depending on the dataset used. 

Systematic uncertainties on the type of hadrons produced and their kinematics are not currently used in the analysis, but are considered for the future as they could matter for the statistical separation of neutrino and anti-neutrino interactions. The different types of pions leave different signatures in the detector, and comparisons between interaction generators show significant differences in the fractions of each type of pion among the hadrons produced in low W DIS interactions. The kinematics of the produced hadrons could matter as well, as in water Cerenkov detector only particles with a high enough momentum (and photons) will appear as rings, particles below the threshold would either not be detected, or could appear as Michel electrons in the case of charged pions and muons.

\newpage

\section{Resonant and Non-resonant Pion Production with W $\ge$ $\Delta(1232)$}
\subsection[Dynamical Coupled-channels Approach to Resonance Region beyond $\Delta(1232)$ - \\Satoshi Nakamura]{Dynamical Coupled-channels Approach to Resonance Region beyond $\Delta(1232)$ - \\Satoshi Nakamura \label{sec:4A}  
[\href{https://indico.cern.ch/event/727283/contributions/3160945/attachments/1732798/2801299/s.nakamura_nustec.pdf}{Presentation}]}

In this presentation, I discuss difficult problems in developing
a neutrino-nucleon reaction model in the resonance region beyond
$\Delta(1232)$ by showing differences between in and above the
$\Delta(1232)$ regions.
I then discuss, as a promising approach,
our dynamical coupled-channels model
for the neutrino-induced reactions beyond $\Delta(1232)$.

In page 3, I show neutrino flux used in ongoing and upcoming neutrino
oscillation experiments.
In the figure on the right, $W$ and $Q^2$ region covered by the neutrino
interactions at a given neutrino energy is displayed.
By combining the two information, we can see that
the neutrino interactions
in the resonance region beyond $\Delta(1232)$ (1.4 \ltap $W$\ltap 2~GeV)
is the main processes utilized in 
some neutrino oscillation experiments like DUNE.

In pages 4-12, I discuss why
neutrino interactions in the resonance region beyond $\Delta(1232)$
are much more difficult to understand than in the $\Delta(1232)$ region.
Main differences between the two energy regions are summarized in the
table shown in page 11.

In pages 13-15, I discuss how one can generally develop a model for
neutrino-nucleon reactions in the resonance region beyond $\Delta(1232)$.
We need matrix elements of the vector and axial currents.
To constrain the vector current matrix elements including the $Q^2$-dependence,
we can use a good amount of electron-induced reaction data such as
single pion productions and inclusive process.
We need both proton and neutron target data to make the isospin
separation of the vector current.
The axial current matrix element is more difficult to control because we
do not have useful data in the resonance region beyond $\Delta(1232)$.
Therefore, we need a theoretical guiding principle which is the PCAC
relation. 
By constructing both pion-nucleon interaction and axial current consistently with
the PCAC relation, we can control not only the magnitude but also 
relative phases among the axial current matrix elements at $Q^2\sim 0$.
In the most previous models, on the other hand, the decay widths of
nucleon resonances are related to the absolute strengths of the axial
matrix elements through the PCAC relation;
the relative phases cannot be uniquely fixed in this way.
Regarding the $Q^2$-dependence of the axial matrix elements,
we neither have useful data nor a theoretical guiding principle.
Therefore it is common to assume the dipole form of the 
$Q^2$-dependence.

In page 16,
I emphasized that our dynamical coupled-channels (DCC) approach~\cite{dcc_nu,knls13,knls16}
follows the strategy, discussed in pages 13-15, to tackle the neutrino interactions 
in the resonance region beyond $\Delta(1232)$.

In pages 17-20,
I discuss theoretical framework of the DCC model.
All observables in reactions are calculated by solving the
coupled-channels scattering equation in which all the relevant coupled
channels are taken into account.
The reactions are driven by meson-exchange non-resonant mechanisms and
bare $N^*$ excitation mechanisms; the latter is dressed by meson cloud
to form a resonance.

In pages 21-28,
I present results of our analysis of reaction data based on the DCC
model.
We first analyzed $\pi N, \gamma N\to \pi N, \eta N, K\Lambda, K\Sigma$
data which include $\sim 27,000$ data points~\cite{knls13,knls16}
After this analysis, we extend the model to finite $Q^2$ region by
analyzing electron scattering data~\cite{dcc_nu}.
I present some selected results to
show the quality of the fits achieved with the DCC model.
Basically, the data are well reproduced.
This reasonable description of the available data is a good basis with
which we can proceed to the neutrino-induced reactions.

In pages 29-35,
I present results of neutrino cross sections predicted by the DCC model.
In the region above the $\Delta(1232)$,
non-resonant and higher
resonances are comparably important.
Two-pion productions are important final states, and the DCC model
is the first model to describe these processes in this energy region.
While the single pion productions are dominated by the $\Delta(1232)$
excitation mechanism,
the two pion productions are mainly from the second and third resonance
regions.

In pages 36-39,
I discuss neutrino cross sections at $Q^2\sim 0$.
These cross sections are not from neutrino experiments, but from the
pion-nucleon cross section via the PCAC relation.
Because there is little experimental information to constrain models for
the resonance region above $\Delta(1232)$, I strongly suggest to use
this PCAC-based neutrino cross sections to constrain the models.
Our DCC model, by construction, well reproduce this data, because we
developed both pion-nucleon and axial current consistently with the PCAC
relation.
On the other hand, the other available models such as the Rein-Sehgal
model~\cite{RS} and the LPP model~\cite{LPP} do not well reproduce this
data because they did not consider the consistency with the
pion-nucleon interactions.

In pages 40-42,
I discuss one of the remaining issues to be addressed.
A model for the resonance region above $\Delta(1232)$ should smoothly
connect to the DIS region where the parton distribution functions (PDF)
describe the processes.
For the electromagnetic sector, this connection is well achieved with
the DCC model.
For the neutrino processes, however, the DCC model does not match well
with the PDF at the boundary between the resonance and DIS regions.
This problem needs a remedy in near future.

In page 43-44,
I make a summary.

\subsection[Single Pion Production in Neutrino Interactions – Minoo Kabirnezhad]{Single Pion Production in Neutrino Interactions – Minoo Kabirnezhad \label{sec:4B} 
[\href{https://indico.cern.ch/event/727283/contributions/3102162/attachments/1732773/2801251/Minoo_NuSTEC18.pdf}{Presentation}]}
Models of SPP cross section processes are required to accurately predict the number and topology of observed charged-current (CC) neutrino interactions, and to estimate the dominant source of neutral-current (NC) backgrounds, where a charged (neutral) pion is confused for a final-state muon (electron). These experiments make use of nuclear targets. The foundation of neutrino-nucleus interaction models are neutrino-nucleon reaction processes like the one described in MK model \cite{MK}.\\
Single pion production (SPP) from a single nucleon occurs
when the exchange boson has the requisite four-momentum
to excite the target nucleon to a resonance state which
promptly decays to produce a final-state pion (resonant
interaction) or to create a pion at the interaction vertex
which is called nonresonant interaction.\\
The SPP processes have been modeled in the $\Delta$ resonance region ($W<1.4~\text{GeV}$, where $W$ is invariant mass) \cite{Adler,HNV,Sato}, and updated to include more isospin $\frac{1}{2}$ resonance states \cite{Fogli,Alam}. However, models for neutrino interaction generators such as NEUT (the primary neutrino interaction generator used by the T2K experiment) \cite{NEUT} require that all resonances up to $W=2~\text{GeV}$ be included to accurately predict neutrino interaction rates. \\
The Rein and Sehgal (RS) model \cite{RS} does include these higher resonances,
but does not include a reliable model for nonresonant processes and related interference terms, and also neglects lepton mass effects.
NEUT and GENIE \cite{GENIE} use the RS model for SPP by default, although they have made minor tweaks and improvements to their implementations 
like NEUT includes charged lepton masses \cite{BS} and a new form factor \cite{GS} (page 4).
In a later paper~\cite{Rein} Rein suggests how to coherently include the helicity amplitudes of the nonresonant contribution (three Born diagrams) to the helicity amplitudes of the original RS model. This update still neglects lepton mass effects (page 5).\\
In the MK model, we improve upon the ideas put forth by Rein by incorporating the nonresonant interactions introduced by Hernandez, Nieves, and Valverde (the HNV model) \cite{HNV}.
The previously neglected lepton mass effects, as well as several other features that make this model suitable for neutrino generators, are also included.
The resulting model has a full kinematic description of the final state particles, including pion angles, for CC and NC
neutrino-nucleon and antineutrino-nucleon interactions (page 1 and 12).\\
It is important to notice that RS model is not a full kinematic model and the output of this model is $d\sigma/dWdQ^2$ where it does not include the pion angles. Page 10 and 11 describe MK model as a full kinematic cross section where the output of the model is $d\sigma/dWdQ^2d\theta_{\pi} d\phi_{\pi}$. 
The main effects of full kinematic cross section and nonresonant interaction appears in pion angles which is shown on page 17.\\
On slides 13-19  the MK mode predictions is compared with NEUT 5.3.6 prediction and bubble chamber data. \\
On slides 20-22 similar comparison is done with T2k and MINERvA data with nuclear targets.\\
On slides 23-24 The MK model prediction for differential cross section is compared with inclusive electron scattering data with hydrogen target. Multi pions can contributes at higher W and energy transferred ($q^{0}$), and this is the main reason for data/ MK model discrepancy at higher W. MK model can also predict the helicity amplitudes of resonances (as it is described on page 25), and MK model prediction for $\Delta$ resonance is compared with MAID data on page 26 where nonresonant effects improve the data/model agreements. As we expected the MK model underestimate data for $S_{1/2}$ since RS model does not predict it for $\Delta$ resonance, and nonresonant background can only contribute in the MK prediction (red curve). Slides 27 shows that MAID analysis is different than JLab analysis for higher resonances.

\subsection[Status of Neutrino-Nucleus Data. Emphasis on Resonances above the $P_{33}(1232)$ – \\Steve Dytman]{Status of Neutrino-Nucleus Data. Emphasis on Resonances above the $P_{33}(1232)$ – \\Steve Dytman \label{sec:4C}  
[\href{https://indico.cern.ch/event/727283/contributions/3102164/attachments/1732817/2801337/sisdata-dytman-2018.pdf}{Presentation}]}

Very little data have been reported for the hadronic content of neutrino interactions in the SIS or DIS regions.  The best data is for H or D targets coming from bubble chamber experiments of many years ago.  The $W$ spectra provide interesting insights into the resonance contributions for various isospin channels.  This is the basis for many existing models.

The only data in this $W$ range for nuclei comes from MINERvA~\cite{McGivern:2016bwh}.  They report $\pi^\pm$ production for $\nu_\mu CH$ and $\pi^0$ production from $\overline{\nu}_\mu CH$ for the NuMI low energy flux.  Since the signal is given as $W<1.8$ GeV, the $\Delta$(1232) excitation is included.  The corresponding $W<1.4$ GeV data are given in the MINERvA data repository~\cite{minerva_datarelease}.

These data provide little specific information beyond the moderate agreement with GENIE v2.8.4.  It should be noted that more modern versions of GENIE are in better agreement.

\newpage

\section{The Transition from SIS to DIS}
\subsection[Quark-Hadron Duality in Neutrino Nucleon Scattering and Analogies in Nuclei - \\Manny Paschos]{Quark-Hadron Duality in Neutrino Nucleon Scattering and Analogies in Nuclei - \\Manny Paschos \label{sec:5A} 
[\href{https://indico.cern.ch/event/727283/contributions/3102167/attachments/1731811/2922374/Paschos-Quark-_Hadron_Duality.pdf}{Presentation}]}

In this talk I include two parts. The first one is historical describing how duality was introduced.  The second describes the progress that has taken place.

In 1968 the SLAC-MIT experimental group presented their data on DIS for moderate values of $Q^2$ where the cross section was large and the resonances were very evident~\cite{ref1}. At the same time Bjorken proposed the scaling of the data~\cite{ref2}. The experimental results created a debate about whether scaling was really observed.  I discussed the experimental results with many colleagues who questioned scaling, pointing that several experimental points were outside the scaling curve.

There were several proposals trying to explain the data. One was based on vector meson dominance~\cite{ref3}; another, the parton model, introduced the scattering of the virtual photon on elementary constituents whose kinematics produced scaling~\cite{ref4,ref5}. The question emerged how the resonances are related to the scaling curve in these models.
 
Studying the curve in detail we were motivated to identify the partons with quarks~\cite{ref5}. We paid attention on the area under the scaling curve $F_2(x)$ which was relatively small, suggesting fractional charges. At soft scattering resonances are formed, but when the interactions become violent, at larger values of $Q^2$, the role of the resonances diminishes and the quark content becomes evident. In addition, as the scattered quarks fly out of the proton they recombine with other quarks and convert again to hadrons (final state interactions).

Bloom and Gilman suggested~\cite{ref6,ref7} that the resonances when plotted in the variable

\begin{equation}
\omega'=1+\frac{W^2}{Q^2}=\omega+\frac{M^2}{Q^2}\,\,\,\,\,\,\text{with}\,\,\,\,\,\,\omega=\frac{2M\nu}{Q^2}\nonumber
\end{equation}
will slide down the scaling curve as $Q^2$ increases and they will eventually  disappear. This is more evident when diffraction is subtracted from the structure functions. The disappearance of the resonances is a dynamic effect closely related to the emergence of the continuum. The resonances are an intrinsic part of the scaling phenomenon for $\nu W_2$. In two articles they analyzed data on electro-production and suggested several properties for neutrino--nucleon scattering. 
Later on the production of baryon resonances induced by neutrinos was studied including four or more resonances~\cite{ref8,ref9}.  It was natural to extend the studies and check the expectations for duality.   Several properties were verified, especially the precise analyses by the groups at Jefferson Lab. 

\begin{enumerate}
\item The agreement was better for the average over protons and neutrons.

\item Duality works better when diffraction is absent as is the case for the function $xF_3(x)$. 

\item The average over the resonances is correlated to the valence contribution. 
\end{enumerate}

The  results are emphasized in several of the presentations at this Workshop,  (see \href{https://indico.cern.ch/event/727283/contributions/3102123/attachments/1731984/2799636/Keppel_nuSTEC.pdf}{A} and  \href{https://indico.cern.ch/event/727283/contributions/3102166/attachments/1732853/2801414/duality_nustec_christy.pdf}{B}).

\vskip 0.5cm
{\bf Application on nuclei}

Nuclei are bound states of protons and neutrons which are their partons (constituents). The resonances on nuclear targets will be their excited states, but for moderate values of $Q^2$ and $\nu$ the nuclei break up. Thus the scaling limit for nuclei is a return to scattering on bound protons and neutrons, modified perhaps by higher twist operators. For this reason each channel, like quasi-elastic scattering or the production of the $\Delta$-resonance, is analyzed with a nuclear model in order to extract the cross sections for bound "protons" and "neutrons" including nuclear corrections.
\begin{itemize}
\item The scaling limit for isoscalar nucleons is practically the same for free and bound nucleons.
\item Differences come for resonances, especially from the nuclear model of FSI and/or Fermi motion.
\item We must include bound state effects.
\end{itemize} 

\vskip 0.5cm
{\bf Summary}

\begin{itemize}
\item An average over resonances is intimately related to the large $x$ region of the scaling curve. 
\item Resonances without diffractive scattering are correlated to the valence component of the scaling curve.
\item We have a dual picture for the interactions of currents with proton and neutrons:  one description is with resonances and the other with scaling.  We found a   kinematic relation between them. However, there is not yet an analytic method describing how one description merges into the other.
\item We hope to obtain information how the scattering on confined quarks develops into bound states. For instance, duality dictates relations among channels of electro- and neutrino-production of the Delta and other resonances in order to satisfy the relation
                                
\begin{equation}
F_2(eN)=\frac{5}{18}F_2(\nu N).\nonumber
\end{equation} 

\item For nuclei we must still understand how to extract from data the resonances produced  by neutrinos on bound  "proton" and "neutron",  then compare the results  with the structure functions. New models on this topic are desirable.
\end{itemize}

\subsection[Higher Twist and Duality in the SIS/DIS Transition Region– Huma Haider]{Higher Twist and Duality in the SIS/DIS Transition Region– Huma Haider \label{sec:5B}  
[\href{https://indico.cern.ch/event/727283/contributions/3102168/attachments/1732918/2801582/H.Haider-NuSTEC.pdf}{Presentation}]}

Most of the neutrino experiments are being performed in the intermediate energy region (1-5 GeV) and it has been estimated
that for this energy range significant number of events will contribute from the deep inelastic region as well as from the duality region~\cite{Alvarez-Ruso:2017oui}.
 Therefore, it becomes important to develop a better theoretical 
understanding in this energy region. Phenomenologically, in the case of DIS process the nuclear medium effects 
for the electromagnetic and weak interactions are found to be the same by  Paukkunen and Salgado, while nCTEQ, Jlab, and Kumano et al.  find them to be different. 
Since for the weak interaction both the vector and axial vector currents 
contribute unlike the case of electromagnetic interaction, therefore, the nuclear medium effects in the case of weak interaction should be different from the medium effects in electromagnetic interaction. Hence,
it is required to properly understand the free nucleon structure functions before studying their modifications in the nuclear medium.

In this presentation, I have discussed the free nucleon structure functions for the charged lepton-nucleon as well as neutrino-nucleon DIS processes, in the wide range of $x$ and $Q^2$. The numerical evolutions have been performed
at next-to-the leading order (NLO) independently for $F_{1}(x,Q^2)$ and $F_{2}(x,Q^2)$, i.e. without using the Callan-Gross relation ($2 x F_1(x) = F_{2}(x)$). 
The nucleon structure functions
have been evaluated with the target mass correction (TMC) effect incorporated by using the operator product expansion approach and the dynamic higher twist (HT) effect which
is included using the renormalon method(also checked for phenomenological approach available in literature). The effects of TMC and HT are important in the kinematic region of high $x$ and moderate $Q^2$.  We have obtained 
the results for the ratio of longitudinal to transverse structure functions, i.e. $R(x,Q^2)=\frac{F_{L}(x,Q^2)}{2 x F_{1}(x,Q^2)}$
and compared them with the results of most widely used phenomenological parameterization given by Whitlow et al.~\cite{Whitlow:1990gk,Whitlow:1991uw} 
as well as with the available experimental data~\cite{Whitlow:1990gk,Whitlow:1991uw,Aubert:1985fx,Benvenuti:1989fm,Seligman:1997mc}. Furthermore, we have also discussed the effect of different cuts on the center of mass (COM) energy $W$.
 
Slides(3-7):DIS kinematics, differential scattering cross section and form of nucleon structure function in terms of nucleon parton distribution functions have been discussed briefly.

Slides(8-13): I have discussed in brief the QCD evolution of partons(NLO), effect of target mass(TMC) and multi-parton correlation(twist-4 HT). I have presented the expressions with these effects, used in numerical calculation of structure functions.
We tried different approaches to take into account  twist-4(shown two approaches here) and TMC(shown only one) corrections, and found no significant difference in the values of structure functions. 

Slide(14): In this slide it is important to note that Callan Gross is not valid for a wide range of $x$ and $Q^2$.

Slides(15-20): In  GENIE,  Whitlow's parametrization($R(x,Q^2)$) has been used for the evaluation of $F_{1}(x,Q^2)$ structure function. We have compared our numerical results of  $F_{1}(x,Q^2)$ (where NLO, TMC, twist-4 corrections have been incorporated) with the Whitlow paramertization. We have found that results in the case of electromagnetic interaction obtained with the present prescription are in fair agreement with the results obtained using  Whitlow parametrization. We must point out that our results obtained for the weak interaction case is slightly different from the electromagnetic case.

Slide(21): We find that nucleon structure function  $F_{2}(x,Q^2)$ do not overlap for EM and Weak processes.

Slides(22-27): We have discussed about the $W$ and $Q^2$ dependences on the EM structure functions $F_{1}(x,Q^2)$ and $F_{2}(x,Q^2)$ which may be useful to understand the transition of resonace to DIS region. Also, presented the curve for $F_{3}(x,Q^2)$ where we observe that inclusion of HT make the results in good agreement with the available IHEP JINR data.

Slide(27-28): We have shown the results for the ratio $R(x,Q^2)$ vs $x$  at $Q^2=7~GeV^2$ and $R(x,Q^2)$ vs $Q^2$  for several values of $x$, for the charged lepton-nucleon as well as the neutrino-nucleon deep inelastic scattering processes. We find that the results for the ratio obtained in the case of electromagnetic interaction
is different than what has been found out for weak interaction. The numerical results are compared with the experimental data of SLAC, BCDMS, EMC and CCFR~\cite{Whitlow:1990gk,Whitlow:1991uw,Aubert:1985fx,Benvenuti:1989fm,Seligman:1997mc} as well as 
with the phenomenological parameterization of Whitlow et al.~\cite{Whitlow:1990gk,Whitlow:1991uw}. 

\subsection[Chiral Field and Regge theory in the Transition Region. Pion Production in a Hybrid Model - Natalie Jachowicz]{Chiral Field and Regge theory in the Transition Region. Pion Production in a Hybrid Model – Natalie Jachowicz \label{sec:5C}  
[\href{https://indico.cern.ch/event/727283/contributions/3102169/attachments/1733088/2801941/jachowicz.pdf}{Presentation}]}

In this talk, we present the formalism developed in the Ghent group for the description of neutrino-induced 1-pion production off the nucleon and nucleus~\cite{piT2k,pinucleus,pinucleon}.  The model was designed to be valid over a  broad kinematic range.
To this end, we combine a 'low energy' model valid in the resonance region (W$\lessapprox$ 1.4 GeV)  with a 'high energy' model,  exploiting the advantages of an approach based on Regge phenomenology to avoid the problematic behavior low energy approaches exhibit at larger invariant mass.

With this model we aim at mending the non-physical behavior of the resonant and background
diagrams in the high-energy regime. The starting point of our framework is a low-energy model
that contains the s- and u-channel diagrams of the P33 (1232) (delta), D13 (1520), S11 (1535), and
P11(1440) resonances and the tree-level background terms derived from chiral perturbation
theory (ChPT) for the $\pi$N system. The resonant amplitudes are regularized by Gaussian 
form factors in order to retain the correct amplitude when s(u)  $\approx M_{res}$ , meanwhile
eliminating the unphysical contributions far away from the resonance peak.
Taking into account higher order diagrams quickly becomes unfeasible. Therefore, in our model
the high-energy behavior is described using a Regge-based approach, where the t-channel
Feynman propagators from the background terms are replaced by the corresponding Regge
trajectories. The low- and high-energy models are combined in a phenomenological way
in a hybrid model that can be used over a broad kinematic region spanned by current and future
experiments.
Next to this applicability to high energy processes, the main strength of our approach is the
treatment of the nuclear matrix element. While most other approaches are restricted either to
single pion production on free nucleons, or are implemented in a nucleus through a Fermi gas model, we treat the
initial  wave functions in the relativistic mean field (RMF) framework which has
proven its merits by providing an excellent description of single nucleon knockout processes.  This can be extended to include RMF final-state wave functions

Slides 3-6 provide a general introduction to the low-energy model on the nucleon and show the limits of the validity of this formalism in a confrontation with data.\\
Slides 7-11 provide the basic ingredients of our high energy description and compare to electron scattering data.\\
Slides 12-14 discuss the implementation of the axial current contributions in the model and show results for the high energy model.\\
Slides 15 and 16 show how the low and high energy approaches are merged into the Hybrid Model in a phenomenological way and present results in a comparison with ANL and BNL data and NuWro predictions.\\
Slides 27-29 explain how the model is implemented in the nucleus and confront our predictions with various MiniBooNE, T2K and MINERvA data.  Moreover, the influence of final-state interactions, not included in the model, is investigated by comparing our predictions with NuWro simulations for full events and events constrained to 1$\pi$1$N$ final states, both with and without FSI.

\newpage

\section{Nuclear Modifications of Structure Functions and Nuclear Parton Distribution Functions}
\subsection[Nuclear Medium Effects in EM and Weak Structure Functions – M. Sajjad Athar]{Nuclear Medium Effects in EM and Weak Structure Functions – M. Sajjad Athar \label{sec:6A} 
[\href{https://indico.cern.ch/event/727283/contributions/3160954/attachments/1732816/2801336/m-sajjad-athar-nustec2018.pdf}{Presentation}]}

In this presentation on the “Nuclear Medium Effects in EM and Weak Structure
Functions” I spoke about the theoretical calculations recently performed by Aligarh-Valencia group to understand nuclear medium effects in the 
 electromagnetic (using charged lepton beam)  and weak(using neutrino/antineutrino) interaction induced deep inelastic reactions on various nuclear targets. 
 I introduced the topic with special mention on the present “Phenomenological” and “Theoretical” efforts going on world wide. 

Neutrino has importance over charged-lepton for having the ability to interact with the particular quark flavors which would
help to understand the partons distribution inside the nucleon. A precise determination of weak structure functions ($F_{iA}^{WI}(x,Q^2$); $i=1,2,3,L$) is required. 
 The weak interaction has additional contribution from axial part and therefore
 medium modifications for weak structure functions may be different from the electromagnetic structure function~\cite{Haider:2016zrk}.\\ 

The various nuclear medium effects which we have considered are
\begin{itemize}
\item  Fermi motion
\item  Pauli blocking
\item  Nucleon correlations
\item  Mesonic contributions
\item  Shadowing and Antishadowing
\end{itemize}

We observe that : 
\begin{enumerate}
\item the nuclear medium effects are different in the weak structure functions from the electromagnetic case.
This difference is $x$ as well as $Q^2$ dependent. This difference increases with the increase in mass number( up to 10-12$\%$ for heavy nuclei like lead).
\item  the nuclear medium effects are different in the different regions of Bjorken variable $x$ as well as $Q^2$
\item  the nuclear medium effects increases with the mass number.
\item  the nonisoscalarity effect is important particularly in heavy nuclei where the variations(nonisoscalarity vs isoscalarity) are as large as 10-12$\%$. 
Moreover, these variations are
 $x$ and $Q^2$ dependent. 
\item the mesonic contributions are important in the mid-region of $x$.
\end{enumerate}

We have
~\cite{SajjadAthar:2007bz,SajjadAthar:2009cr,Haider:2011qs,Haider:2012nf, Haider:2012ic, Athar:2013gya, Haider:2015vea, Haider:2014iia,Haider:2016tev,Haider:2016zrk} 
 studied nuclear medium effects in the structure functions in a microscopic model using 
\begin{enumerate}[label=\Roman*]
\item: Relativistic nucleon spectral function to describe target nucleon momentum distribution incorporating Fermi motion, binding energy effects and nucleon 
correlations in a field theoretical model. The spectral function that describes the energy and momentum distribution of the nucleons in nuclei is obtained by
using the Lehmann’s representation for the relativistic nucleon propagator and nuclear many body theory is used to calculate it for an interacting Fermi sea 
in nuclear matter~\cite{FernandezdeCordoba:1991wf}.
A local density approximation is then applied to translate these results to a finite nucleus. Nuclear information like Binding energy, Fermi motion, nucleon 
correlations, is contained in the spectral function.\\

\item: The contributions of the pion and rho meson clouds are important in the mid region of $x$ which arises due to the presence of strong attractive nature of
nucleon-nucleon interactions, which in turn leads to an increase in the interaction probability of virtual boson with the meson clouds. To take into account the 
mesonic effect, microscopic approach has been used by making use of 
the imaginary part of the meson propagators instead of spectral function in a many body field theoretical approach following Refs.~\cite{Marco:1995vb,GarciaRecio:1994cn}.\\

\item: Non-isoscalarity corrections are taken properly into account by normalizing the spectral function independently for the proton and neutron targets in a nucleus.\\

\item: Shadowing effect has been taken into account following the works of Kulagin and Petti~\cite{Petti2}.\\

\item: These calculations are performed without assuming Bjorken limit. Evolution of each nucleon structure functions $F_1(x;Q^2)$, $F_2(x;Q^2)$ and $F_3(x;Q^2)$, 
which are inputs in calculating nuclear structure functions, are done independently. For the nucleon PDFs CTEQ6.6 parameterization is used.
Slide-27 shows the expressions of nuclear structure functions $F_1^A(x;Q^2)$, $F_2^A(x;Q^2)$ and $F_3^A(x;Q^2)$.\\
\end{enumerate}

\begin{itemize}

\item Slide-30 and 31 show the  expressions of mesonic structure functions.

\item Slide-32 show the dependence of pion parton distribution functions phenomenologically obtained by various groups like CTEQ5L, GRV, Conway, MRST98, SMRS, on the
 electromagnetic structure function $F_2(x,Q^2)$.

\item Slide-33 LHS: In this figure we show the results for the ratio of electromagnetic structure functions in Beryllium to Deuterium. The dashed line has been calculated with the spectral function only with TMC and the solid line corresponds to the full model, including the meson cloud contributions, shadowing and TMC. We show explicitly the effect of  shadowing. It reduces the structure function ratio by around 1\%  at $x\sim0.3$ and even less for higher $x$. We have found that TMC has a really minor effect in the ratio for these $x$ values (less than 1\% at $x\sim0.6$ and even smaller for lower $x$ values). 
 Therefore, the difference between the results obtained using spectral function(base curve) and the full one comes basically from the $\pi$ and $\rho$ contributions that play an important role.  The size of the  rho meson correction is about half that of the pion. We find that the full model agrees quite well with data both in slope and the size of the ratio. 

RHS: A good agreement with data is also obtained for Carbon as shown in the figure on RHS. The slope and size of the nuclear effects are similar to the Beryllium case. We have also shown here the effect of varying the parameter associated with spin-isospin interaction in the expression of meson self energies. We find that a 20$\%$ variation in the $\Lambda$'s, results in a 2-3$\%$ change in the ratio.

\item Slide-34 This figure shows the results for the weak structure function $F_2(x,Q^2)$ in iron. The effect of  varying the parameter associated with spin-isospin interaction in the expression of meson self energies has also been shown.

\item Slides-36-38 show the results for the electromagnetic structure functions in various nuclei at several $Q^2$  and the results are compared with Jlab, NMC and SLAC data.

\item Slide-39 show the results for the weak structure functions $F_2$ in iron at several $Q^2$ and the results are compared with the experimental results from CCFR, CDHSW, NuTeV as well as the phenomenological fit of Tzanov.

\item Slide-40 show the results for the electromagnetic and weak structure functions $F_2$ and results are compared with CCFR, CDHSW, NuTeV and EMC data. 
 It may be observed that the results for the weak structure function is different from the electromagnetic structure function specially at low $x$. This difference 
 decreases with the increase in $Q^2$(not shown here).

\item Slide-41 show the results for the weak structure function $F_3$ in iron with the spectral function and in the full model. 
 The results are also shown for the free nucleon case as well as these are compared with the experimental data from  CDHSW and NuTeV collaborations.

\item  To observe the variation of nuclear medium effects in the electromagnetic and weak interaction processes, we obtain the results for the ratio $R^\prime=\frac{\frac{5}{18} F_{2A}^{WI}(x,Q^2)}{F_{2A}^{EM}(x,Q^2)}$ in  various nuclei like 
 $^{12}$C, $^{27}$Al, $^{40}$Ca, $^{56}$Fe, $^{63}$Cu, $^{118}$Sn and $^{208}$Pb at $Q^2=6$ and $20$ GeV$^2$. 
 These results are obtained assuming the targets to be nonisoscalar wherever applicable. One may observe that for heavier nuclear targets like  $^{118}$Sn and $^{208}$Pb, the nonisoscalarity effect is larger.  This shows that the difference in charm and strange quark distributions could be
  significant in heavy nuclei. 
\end{itemize}

\subsection[Nuclear Medium Effects on the Structure Functions – Sergey Kulagin]{Nuclear Medium Effects on the Structure Functions – Sergey Kulagin \label{sec:6B}  
[\href{https://indico.cern.ch/event/727283/contributions/3160955/attachments/1733266/2802262/kulagin-nustec2018.pdf}{Presentation}]}

In this presentation, on one side, I summarize experimental data on the nuclear EMC effect on the parton level
obtained in deep-inelastic and Drell-Yan experiments with nuclei and, on the other side, I discuss recent advances
in the development of a microscopic model of nuclear structure functions and the parton distributions,
which is based on solid physics mechanisms responsible for nuclear corrections and shows excellent performance
in comparison with data. 

The data on the nuclear EMC effect is summarized in page 3 to 6 of my presentation (from now on the references to pages refer to my slides).
Page 4 shows a matrix plot of the ratio of structure function $F_2$ of a heavy nucleus to deuterium, $R(A/D)=F_2^A/F_2^D$, as a function of Bjorken $x$ for a number of nuclei from ${}^4$He to ${}^{208}$Pb. The data shown are from the measurements at CERN (EMC~\cite{Ashman:1992kv}, NMC~\cite{Amaudruz:1995tq,Arneodo:1996rv} and BCDMS~\cite{Bari:1985ga}), SLAC E139~\cite{Gomez:1993ri}, DESY HERMES~\cite{Ackerstaff:1999ac}, FNAL E665~\cite{Adams:1995is}, JLab E03103~\cite{Seely:2009gt}. Page 5 shows data for $^3$He from HERMES~\cite{Ackerstaff:1999ac} in comparison with data of JLab E03103 experiment~\cite{Seely:2009gt}, and pg.6 presents the results of JLab BoNuS~\cite{Tkachenko:2014byy,Griffioen:2015hxa} and SLAC E139~\cite{Gomez:1993ri} for deuterium.

In page 7 to 13 I briefly review a number of mechanisms of nuclear corrections including the effects due to energy-momentum distribution of bound nucleons (Fermi motion and nuclear binding) developed in Refs.~\cite{Akulinichev:1985ij,Kulagin:1989mu,Kulagin:1994fz,Kulagin:2004ie},
off-shell correction~\cite{Kulagin:1994fz,Kulagin:2004ie}, nuclear meson-exchange currents in DIS, as well as nuclear shadowing.

Page 14 and 15 summarize different basic ingredients of a microscopic model of nuclear DIS structure functions of Refs.\cite{Kulagin:2004ie,Kulagin:2007ju,Kulagin:2010gd,Kulagin:2014vsa}. Page 16 gives a summary of phenomenological determination of the function $\delta f(x)$ responsible for off-shell correction from two independent  studies~\cite{Kulagin:2004ie,Alekhin:2017fpf}. On pg.17 to 18 I discuss physics interpretation of this function in terms of the nucleon scale parameters. Using the model of Ref.\cite{Kulagin:1994fz} I argue that the observed shape and crossover of $\delta f(x)$ indicate the increase of the bound nucleon core radius in nuclear environment (about 10\% for heavy nuclei such as  iron and 2\% for deuterium). On pg.19 I discuss a relation between nuclear shadowing and off-shell correction using the sum rule for the valence quark number. Page 20 to 23 summarize the results of Refs.\cite{Kulagin:2004ie,Kulagin:2010gd} on the nuclear ratios in comparison with data.
See also a table on pg.34 in the Backup section summarizing the values $\chi^2$ computed between our model predictions and data from various experiments, pg.41 which
illustrate different nuclear corrections on the example of $^{197}$Au, pg.42 to 45 which  summarize the analysis of the EMC effect in the deuterium and $^3$He nuclei.

On pg. 24 to 25 we compare $F_2$ computed using two different fits: DIS analysis based on QCD NNLO PDF fit of Ref.\cite{Alekhin:2006zm,Alekhin:2007fh} 
and a fit of Ref.\cite{Christy:2007ve,Bosted:2007xd} which includes a parameterization of the nucleon resonance contributions as well as nonresonant background. A matrix plot on pg.24 shows such a comparison for the fixed values $Q^2=1,\ 2,\ 4\:\mathrm{GeV}^2$ for the proton (first column), the neutron (2nd column) and the deuteron (3rd column) as a function of $W^2$. We observe a nice agreement between the DIS and the resonance analyses for the proton in the overlap region, while some mismatch in the normalization is present for the neutron. The latter could be related to a different treatment of nuclear effects in the deuteron in Refs.\cite{Bosted:2007xd} and \cite{Alekhin:2007fh}
in the extraction of the neutron data.

A duality relation between the resonance and the partonic descriptions is discussed on pg.25.

In order to quantitatively address the domain of ``shallow'' inelastic region we discuss a hybrid model combining the resonance and the partonic descriptions. This model, which utilizes the results of Refs.\cite{Christy:2007ve} and \cite{Alekhin:2007fh}, is  summarized on pg.26 to 27. The performance of this model in comparison with the proton and deuteron data is illustrated by the figures on pg.28 to 29. The figures on pg.28 show the proton and the deuteron $F_2$ computed as a function of $W^2$ for the fixed values $Q^2=1.025,\ 2.275,\ 2.525,\ 3.525\:\mathrm{GeV}^2$ in comparison with data from JLab~\cite{Osipenko:2003bu,Osipenko:2005gt}, SLAC~\cite{Whitlow:1991uw}, CERN NMC~\cite{Arneodo:1996qe}. For a better visibility the deuteron curves are shifted up by 0.2. Figures on pg.29 show a comparison of our predictions with JLab BoNuS data~\cite{Tkachenko:2014byy}  on the ratio of the neutron and the deuteron structure functions $F_2^n/F_2^D$. Figures on pg.30 illustrate behavior of the ratio $(F_2^p+F_2^n)/F_2^D$ as a function of Bjorken $x$ computed for a few different $Q^2$  using the DIS model (left panel) and the hybrid DIS+RES model (right panel). For more detail and comparison of model predictions with data see Ref.\cite{Kulagin:2018mxb}.

Page 31 summarizes our results for the nuclear effects in $^3$He. Figure in the left panel summarizes the results of the measurement of the ratio $F_2^\mathrm{3He}/F_2^D$ in Refs.\cite{Ackerstaff:1999ac} and \cite{Seely:2009gt} in comparison with our predictions using the DIS model (see also Ref.\cite{Kulagin:2010gd}) and the DIS+RES model. Following Ref.\cite{Kulagin:2010gd} we apply the factor 1.03 to data of Ref.\cite{Seely:2009gt}. This allows us to reconcile the normalization of data points from  Ref.\cite{Ackerstaff:1999ac} and Ref.\cite{Seely:2009gt} in the overlap region. Also the ratio $F_2^n/F_2^p$ extracted from data on $^3\mathrm{He}/D$ of Ref.\cite{Seely:2009gt} becomes consistent with that extracted from the NMC measurement of $F_2^D/F_2^p$~\cite{Kulagin:2010gd}. Note that in the region $0.55<x<0.9$ JLab data correspond to $W<2\ \mathrm{GeV}$ and seems to reflect the resonance behavior of the proton and the neutron structure functions  which is described by our model. This resonance behavior is more pronounced in the ratio $F_2^\mathrm{3He}/(F_2^D+F_2^p)$ which is shown in the right panel. The region $x\to1$ requires special consideration as it involves quasielastic contribution which can be significant for JLab kinematics.

In summary (pg.32), the data on the nuclear EMC effect in the DIS/SIS region can be understood by addressing a number of corrections including the nuclear binding and momentum distribution effects, target-mass and off-shell correction, meson-exchange currents as well as the matter propagation effects of hadronic component of virtual photon. Those nuclear effects result in the corrections relevant in different kinematic domains. In the resonance region  the ratios of nuclear structure functions show a strong oscillating behavior. 

In the Backup section, pg.46 to 57 summarize our results on the nuclear DY process~\cite{Kulagin:2014vsa}, nuclear PDF and application for $W/Z$ boson production in $p+\mathrm{Pb}$ collisions at LHC~\cite{Ru:2016wfx}.

\subsection[nPDFs from $l^{\pm} A$ and $\nu A$ scattering – Aleksander Kusina ]{nPDFs from $l^{\pm} A$ and $\nu A$ scattering – Aleksander Kusina \label{sec:6C} 
[\href{https://indico.cern.ch/event/727283/contributions/3102173/attachments/1733314/2802603/kusina_NuSTEC2018.pdf}{Presentation}]}

The purpose of this presentation was to summarize the results on compatibility
of nuclear parton distribution functions (nPDFs) determined from the charged
lepton deep inelastic scattering (DIS) data and from the neutrino DIS data. I concentrated
on the initial works from the nCTEQ group that point to the incompatibility of
the nPDFs obtained from these two data
types~\cite{Schienbein:2007fs,Schienbein:2009kk,Kovarik:2010uv},
and later mention newer studies addressing this issue.

I start (in slides 2-7) by presenting the basic ideas behind nPDFs and how they
are determined by fitting experimental data in the process of global QCD analysis.
I tried to highlight their direct connection to the QCD factorization theorems~\cite{Ellis:1978ty,Collins:1985ue,Bodwin:1984hc,Qiu:2003cg}
and the fact that the universality of PDFs is a direct consequence of these theorems.
In all PDF analyses the backbone of the fit is provided by the DIS data. The typical
data used in nPDF analyses are the changed lepton (neutral current) DIS data from
fixed-target experiments like NMC, BCDMS or EMC e.g.~\cite{Arneodo:1996ru,Arneodo:1996rv,Bari:1985ga,Ashman:1988bf}.
However, there are also available DIS neutrino (charged current) data from the
NuTeV~\cite{Tzanov:2005kr},
CCFR~\cite{Yang:2000ju},
CHORUS~\cite{Onengut:2005kv},
and
CDHSW~\cite{Berge:1989hr}
experiments.
From the point of view of global PDF analysis the best approach is to use as much data as
possible, covering different processes, as it allows for better determination of distributions
of individual partons (different processes are sensitive to different PDF combinations).
In addition, it allows to test the factorization framework which predicts the same PDFs/nPDFs
regardless of the involved hard process.

In slides 9-12 I presented the main results of the nPDF analysis of ref.~\cite{Schienbein:2007fs}
which extracted nPDFs for iron from the NuTeV and CCFR neutrino data~\cite{Tzanov:2005kr,Goncharov:2001qe}.
Rather surprisingly the analysis found that the obtained nuclear corrections are different than
the ones obtained from the charged lepton data~\cite{Kulagin:2004ie,Hirai:2007sx}.
In particular, featuring a shadowing that is delayed to much smaller $x$ values.
This result triggered a separate nPDF analysis from the nCTEQ group based only on the charged
lepton DIS data and Dell-Yan (DY) data~\cite{Schienbein:2009kk}, which is described in slides
14-17. The obtained nPDFs were in agreement with the earlier works~\cite{Kulagin:2004ie,Hirai:2007sx}
confirming the apparent difference between nuclear modifications determined from the charged
lepton and neutrino DIS data.
Nevertheless, the question reminded whether it would be possible to obtain a compromise fit
including both charged lepton and neutrino DIS data at the same time. This question has been
addressed in ref.~\cite{Kovarik:2010uv}. In slides 19-30 I summarize this analysis in a bit
more detail providing also some additional information.
In slide 19 I start by detailing the experimental data involved in the analysis together with
kinematical selection ensuring that the used data is in the kinematic region where collinear
factorization is valid. In slide 20 the approach to the analysis is explained which consisted
of performing a series of nPDF fits where charged lepton DIS data is supplemented by the neutrino
data with different weights equal to $\{0,1/7,1/4,1/2,1,\infty\}$ ($w=0$ corresponds to a fit
with only charged lepton data and $w=\infty$ to a fit with only neutrino data). One should note
that different values of weights can additionally compensate for quite different numbers of data
points in the neutral and charged current data sets (708 data points in case of charged lepton
data and 3134 in case of neutrino data).
Slide 21 presents the obtained results on the level of the nuclear modification of the $F_2$
structure function showing how the fits with different weights interpolate between the extreme
scenarios. Of course, the obtained fits are not equally good in terms of the $\chi^2/$dof values.
These values are detailed in slide 22 and potential compromise fits are selected.
The next slide shows the results of the fits on the level of nPDFs and compares them with the
current nPDFs from the nCTEQ15 analysis~\cite{Kovarik:2015cma} which features Hessian errors not
available at the time of the analysis of ref.~\cite{Kovarik:2010uv}. One can observe here a
discrepancy between the fits involving the neutrino data and the charged lepton only fits,
in particular, for the best known valence distributions in the region of $x\lesssim0.1$.
This discrepancy also persist when the error band of the nCTEQ15 fit is accounted for (this fit
also does not used the neutrino DIS data). 
The next three slides provide additional discussion and introduce a tolerance criterion that is
used to judge the compatibility of a fit (with a given $\chi^2$) with the best fit on the level of
individual data sets. The 90\% CL and 99\% CL tolerances are determine and visualized in slide 27
for individual fits. These are used to draw conclusions on the fits with different weights which
are summarized in slide 28. Slides 29 and 30 provide conclusions of the whole series of the presented
nCTEQ studies which showed that:
(i) when a weight of neutrino data is increased the description of the NuTeV data gets better but
it is always accompany by a degradation of the description of the charged lepton data;
(ii) study based on statistical tolerance criterion confirmed qualitative observations based on the
$F_2$ ratio plots;
(iii) based on the 90\% CL criterion there is no compromise fit, relaxing the criterion to 99\% CL
the fit with $w=1/2$ becomes marginally acceptable.
One should highlight that the observed tensions are originating mostly from the NuTeV data, and in
particular, are made much more pronounced by using the (unpublished) correlated errors.
The CCFR and CHORUS data are generally compatible with the charged lepton data, however, they also
feature larger uncertainties.

In slides 31-34 I briefly summarize other analyses addressing the above questions.
First, in slide 31 results of ref.~\cite{Kalantarians:2017mkj} are presented. In this
case special care has been taken to work on the level of absolute structure functions instead of
ratios in order to minimize the impact of the deuteron nuclear corrections on the results.
This study also points to inconsistency between the charged lepton and neutrino structure functions
at small $x$ values.
There were also other analyses from the perspective of global nPDF fits which are mentioned in
slides 32-34. In ref.~\cite{deFlorian:2011fp} the authors did not observe any tensions with the
NuTeV data, however, they have used the less precise structure function data and not the data on
the cross-sections.
A more detailed analyses on the level of cross-sections (but using the uncorrelated uncertainties)
were performed by the EPS group~\cite{Paukkunen:2010hb,Paukkunen:2013grz}. They have also observed
certain problems with the NuTeV data which they attributed to the energy dependence of the data.
Also the HKN group showed preliminary results~\cite{Kumano_seminar} exhibiting properties very
similar to what was found in the nCTEQ studies~\cite{Schienbein:2007fs,Schienbein:2009kk,Kovarik:2010uv}.

At the moment it is rather clear that there is some tension between the NuTeV data and the charged
lepton DIS data. However, the question remains whether this tension is really a tension between the
neutrino DIS and charge lepton DIS processes (resulting e.g. from different physical properties of
the charged current and neutral current interactions) or whether there is some problem with the NuTeV
data. Probably to answer this question definitively we will need an independent confirmation from
another neutrino experiment. It view of this it is really unfortunate that not all the data collected
by the NOMAD experiment~\cite{Samoylov:2013xoa} have been published, as they certainly could have
helped to answer this problem.

\newpage

\section{Hadronization in the Nuclear Environment}
\subsection[Hadronization Models in GENIE – Costas Andreopoulos]{Hadronization Models in GENIE – Costas Andreopoulos \label{sec:7A}
[\href{https://indico.cern.ch/event/727283/contributions/3159400/attachments/1733413/2802605/Andreopoulos-GENIE_Hadronization-v1.pdf}{Presentation}]}

In this talk, we presented an overview of the hadronization model in GENIE. Before the intranuclear hadron transport simulation, hadrons can be produced by several models and codes within GENIE. This talk was focused on the models and strategies used in the absence of a model of exclusive production (e.g. single-pion or single-Kaon), when hadronic distributions are not easily calculable. 
Three models were presented in some detail:
\begin{itemize}
    \item An empirical model for Shallow- and Deep-Inelastic Scattering (SIS/DIS), valid for hadronic invariant masses below 3 GeV.
    This empirical model is described in slides 14-39,
    outlining the simulation strategy as well as the key assumptions and experimental 
    data baked into that simulation. Some of the main caveats of this empirical model are 
    summarized in slide 39.
    \item The GENIE interface to the PYTHIA6 Monte Carlo code, employed for the simulation of DIS events in the kinematic space left uncovered by the above empirical model.
    Slides 10-12 outline the specifics of the GENIE/PYTHIA6 interface, highlighting the parameters of the GENIE/PYTHIA6 tune and the oddities arising from the quark / diquark code assignments made by GENIE, which is an area of possible improvement as GENIE migrates to PYTHIA8.     
    \item An empirical model specialized for DIS charm production, 
    which is outlined in slides 40-42
\end{itemize}
Several pieces of data exist for the validation and tuning of hadronization models, 
and we presented GENIE comparisons against:
\begin{itemize}
  \item average charged and neutral particle multiplicity data,
  \item average multiplicity data in the forward and backward hemispheres,
  \item multiplicity dispersion data 
        as function of the multiplicity and hadronic invariant mass W,
  \item multiplicity correlation data 
        (e.g. charged hadrons - $\pi^{0}$ multiplicity correlations),
  \item fragmentation functions,
  \item Feynman x ($x_{F}$) distributions,
  \item transverse momentum ($p_{T}^{2}$) distributions, and
  \item $x_{F}$ - $p_{T}^{2}$ distributions
\end{itemize}

Finally, slides 43-46 present a brief survey of data and models on in-medium effects to hadronization. 

Relevant citations are given throughout the presentation.


\subsection[Hadronization in eA collisions - Kai Gallmeister]{Hadronization in eA collisions - Kai Gallmeister \label{sec:7B}  [\href{https://indico.cern.ch/event/727283/contributions/3160956/attachments/1733393/2802551/Talk2.pdf}{Presentation}]}

The intention of this talk was to bring back into the minds some studies, that were performed already a decade ago with lepton beams.

We recapitulated our main findings from this time, that combining the experimental data obtained by HERMES ($E_e=27.3\,{\rm GeV}$) with that of the EMC collaboration ($E_e=100-200\,{\rm GeV}$), only a {\textbf  linear} increase of the interaction during the formation of the hadrons is compatible with data. This contradicts naive pictures of hadronization and formation times often found in neutrino generators, but is in agreement with some quantum mechanical picture.

GiBUU is the only model, which successfully describes all the HERMES data. At the moment, also successful comparisons to recent CLAS data are on their way. 

The relevant citations are given on the last slide of the presentation.


\subsection[Neutrino Interactions in FLUKA: NUNDIS – Paola Sala]{Neutrino Interactions in FLUKA: NUNDIS – Paola Sala \label{sec:7C} 
[\href{https://indico.cern.ch/event/727283/contributions/3159422/attachments/1733424/2802628/Sala-nundis_gssi_dis.pdf}{Presentation}]}

The FLUKA neutrino event generator, called NUNDIS, is fully embedded in the FLUKA nuclear environment, the same used  for hadronic interactions, photonuclear interactions, muon capture etc.  NUNDIS describes the neutrino-nucleon interactions including  Quasi Elastic, resonance production and Deep Inelastic Scattering. As described in slide 5, QE is based on the Llewellyn Smith formulation, RES is Rein-Sehgal keeping only the Delta contribution. A linear decrease of both DIS and RES cross sections as a function of W removes double counting of pion production.
Slide 6 describes the basic assumption about DIS, that is the composition of two steps, one describing the boson exchange, the other dealing with hadronization. Hadronization is performed with the FLUKA models, again the same as in hadronic interactions, assuming chain universality. 
Slides 7 to 9 give detail of the DIS implementation in terms of parton distribution functions, including recent developments on the extrapolation to low $Q^2$. Slide 10 successfully compares data and NUNDIS on total $\nu_\mu$ CC cross section on nucleons.
Basic ingredients of FLUKA chain hadronization are given in slides 11 and 12. Of relevance for neutrino interactions are the special treatments of low-mass chains, where mass effects become important and are take into account in FLUKA. In addition, for very low mass chains the standard hadronization is replaced by a sort of phase space explosion. This last improvement, originally introduced to improve pion production from low energy hadron-nucleon interactions (slide 13) , has proven to be important for the correct simulation of single-pion production in neutrino interactions. Slide 14 shows this effect: in the plots, we report data (symbols) and simulations (lines)  for total and single-pion cross sections, from $\nu_\mu$ CC interactions on protons and neutrons separately. The various channels composing the total cross section in NUNDIS are also shown. Two results are plotted for the single-pion cross section in FLUKA, before and after the introduction of phase space explosion in hadronization. The agreement with experimental data largely improves when the new treatment is included. This also shows that the DIS contribution has a strong effect on the single-pion channel, which is traditionally associated to resonance production only.   
Slide 15 shows hadronization benchmarks at higher energies, and slide 16 is a nice comparison with data on charm production in neutrino interactions.
The intermediate energy hadronic model of  Fluka is  called PEANUT. 
The reaction mechanism is modelled in PEANUT by explicit intranuclear cascade (INC) smoothly joined to statistical (exciton)
preequilibrium emission (slide 16) 
At the end of the INC and exciton chain, the evaporation  of nucleons and nuclear fragments is performed, following the Weisskopf
treatment.
Competition of fission with evaporation has been implemented, again  within a statistical approach. 
Since the statistical evaporation model becomes less sound in light nuclei,
the  so called Fermi Break-up model  is used instead.
The excitation energy still remaining after (multiple) evaporation is
dissipated via emission of $\gamma$ rays.

The INC proceeds through hadron multiple collisions in a cold Fermi gas formed by bound nucleons.
The hadron-nucleon cross sections used in the calculations are 
the free ones modified by Pauli blocking, 
except for pions and negative kaons that deserve a special
treatment. The Fermi motion is taken
into account, both to compute the interaction cross section, and to produce the
final state particles. 

Secondaries are treated exactly like primary particles, with the only
difference that they start their trajectory already inside the nucleus.
Primary and secondary particles are transported according to their
nuclear mean field and to the Coulomb potential. All particle are
transported along classical trajectories, nevertheless a few relevant
quantum effects are included, such as Pauli Blocking, nucleon antisymmetrization effects and nucleon-nucleon hard-core correlations. 

Binding Energies ($B_{en}$) are obtained from mass tables, depending on
particle type and on the actual composite nucleus, which may differ
from the initial one in case of multiple particle emission. First excited nuclear levels are also included when available.
Relativistic kinematics is applied, with accurate conservation of energy 
and momentum, and with inclusion of the recoil energy and momentum of 
the residual nucleus.

In both stages, INC and exciton, the nucleus is modelled as a sphere
with density given by a symmetrized Woods-Saxon shape for A$>$16, and by a harmonic oscillator shell model for light
isotopes.

A standard position dependent Fermi momentum distribution is 
implemented in PEANUT,  with addition of momentum smearing , as in slide  19

For pions, a nuclear potential has been calculated starting from the standard
pion-nucleus optical potential.
The treatment of pion interactions includes modifications to the $\Delta$ properties in the nuclear medium, following the formulation proposed by Oset et al. (slide 22). Results on pion-induced reactions , and in specific pion absorption, are very good, as shown in slide 23.
 Pion  reactions and absorption are among  the most important  Final State interactions in neutrino induced interactions. As shown in slide 24, FSI drastically change the pion multiplicity and energy spectra. However, existing data from MINERvA and MiniBooNE cannot be simultaneously reproduced, neither by Fluka nor by other models, thus making it difficult to assess the validity of the assumed FSI.

Slide 25 introduces the concept of Formation Zone. It can be understood considering that hadrons are composite objects and that the typical time of strong interactions is of the order of 1~fm. If one thinks about the hadrons emerging from an inelastic interaction, it requires some time to them to ``materialize" and be able to undergo further interactions. Formation Zone has proven to be an essential ingredient to correctly model hadronic interactions. The saturation of the number of medium/low energy secondaries  (mostly reinteraction products)  with increasing projectile energy, as opposed to the increase in the number of “fast” secondaries, calls for a mechanism that  reduces reinteractions. This mechanism has to be energy-dependent, not to spoil results  at lower energies. (slide 26). Other examples are shown in slide 27, and  and example of the effect on neutrino interaction product is in slide 28.
Slides 30 to 34 contain examples of application to real complex experiments, such as Icarus T600 at Gran Sasso.
As started in slide  36,  the implementation of neutrino interactions in Fluka gives promising results, being supported by all the past developments on hadronic interactions. Improvements are still needed, namely the inclusion of coherent scattering and coherent effects at low energies. Comparisons with data shall be extended.
References in the slides 



\subsection[Challenges of Modelling Neutrino induced Shallow Inelastic scattering (SIS) Interactions for Neutrino Oscillation Experiments around 1-10 GeV – Teppei Katori]{Challenges of Modelling Neutrino induced Shallow Inelastic scattering (SIS) Interactions for Neutrino Oscillation Experiments around 1-10 GeV – Teppei Katori \label{sec:7D} 
[\href{https://indico.cern.ch/event/727283/contributions/3174967/attachments/1732255/2806396/TK_XS_nSDIS18.pdf}{Presentation}]}

Although there are successful models of hadronization in collider physics, it is not straightforward to apply them to neutrino experiments. Difficulties in the modeling of neutrino hadronization process lie on the very topics of this workshop. A key element is how to utilize aspects of the hadronization program developed for collider experiments (such as PYTHIA, please see \href{https://indico.cern.ch/event/727283/contributions/3102176/attachments/1733412/2802604/talk20181013.pdf}{presentation} by Stefan Prestel) for low energy neutrino scattering, while incorporating nuclear effects.
\begin{itemize}
    \item[1] Current and future accelerator-based neutrino experiments are interested in measuring hadronization processes with low invariant mass ($W$) which is not the region often studied in collider experiments. 
    \item[2] Current and future accelerator-based neutrino experiments use nuclear targets.
\end{itemize}
In the very low W region ($W<3$~GeV), empirical laws are applied to reproduce neutrino bubble chamber data in simulation. This includes modeling of averaged multiplicity for charged hadrons, modeling of neutral hadron multiplicity, and dispersion relations of them. The AGKY model~\cite{Yang:2009zx} used in GENIE or similar approaches in NEUT are both successful. 

\vskip 0.5cm
{\bf 1. Application of PYTHIA for neutrino oscillation experiments}

The problem is, hadronization laws extracted from neutrino bubble chamber data seem incompatible with the PYTHIA hadronization program.  First of all, neutrino experiments are extending PYTHIA to low W region, say $W<5$~GeV and this introduces problems because PYTHIA is invalid in such low W region. Second, PYTHIA underestimates averaged hadron multiplicity even at the high W region. Because of this, all generators, GENIE, NEUT, NuWro, and GiBUU, predict lower averaged multiplicity than neutrino hadron multiplicity data~\cite{Kuzmin:2013tza}. This second problem is now understood because the default setting of PYTHIA is not suitable for neutrino experiments and Lung string function needs to be tuned for neutrino experiments~\cite{Katori:2014fxa}. Third, dispersion laws extracted from bubble chamber data are much wider than the PYTHIA predicted dispersion law. Because of this, topological cross sections make discontinuity curves with a function of W. Forth, it is hard to model charged hadron and neutral hadron ($\pi^{\circ}$) multiplicity data simultaneously. One known problem is that $\pi^{\circ}$ multiplicity data is usually lower quality because of the difficulty of measurements by bubble chambers. Thus, this moment we cannot conclude that the apparent isospin violation seen between charged hadron and neutral hadron multiplicities are physical or not. 

\vskip 0.5cm
{\bf 2. Hadronization from nuclear target}

Recently, MicroBooNE made the first measurement of charged hadron multiplicity measurement using a liquid argon time projection chamber (LArTPC)~\cite{Adams:2018fud}. This reminds us how to use such data to improve the hadronization simulation for neutrino scatterings. Interpretation of heavy target data is not easy. First, data is a convolution of three effects, primary hadronization, secondary final state interaction (FSI) in the target nuclei, and the tertiary interaction of hadrons in detector media. To study the modeling of hadronization at the primary level, physics of secondary and tertiary processes must be corrected, however, the FSI in target nuclei is in this moment the root of all evil in neutrino interaction physics. We are hoping neutrino production pion data sheds light on this problem. Recently, MINERvA published~\cite{Stowell:2019zsh} their attempt of simultaneous tuning of simulation using the data from 4 channels: $\nu CC1\pi^+$, $\nu CCN\pi^+$, $\nu CC1\pi^\circ$, and $\bar\nu CC1\pi^\circ$. This work tried to tune both primary pion production models and secondary FSI models in the GENIE simulation. The effort is so far not very successful because both cross-section and FSI models used in GENIE do not have enough freedom to remove tensions between each data set. 
\vskip 0.5cm
{\bf Summary}

In summary, we identify four first-order systematic errors on averaged hadron multiplicity.
\begin{itemize}
  \item[(1)] Low W hadronization error, which is not modelled by hadronization programs such as PYTHIA.
  \item[(2)] High W hadronization error, this is an inherent systematic error in hadronization program such as PYTHIA.
  \item[(3)] Charged hadron - neutral hadron ratio error to incorpor ate current data-simulation mismatch.
  \item[(4)] A-dependent hadronization error, because most hadronization process validation data are from limited elements.
\end{itemize}

Currently, GENIE has (1) as re-weighting, meaning systematic errors of low W hadronization is readily available by users. (2) is considered in some analyses in DeepCore~\cite{Aartsen:2017nmd,Aartsen:2019tjl}, and the effect is evaluated to be small. (3) and (4) are not considered by any current experiments, but their impacts might be non-negligible for current and future $\nu_e$ ($\bar\nu_e$) appearance experiments where $\pi^{\circ}$s are main background. Compared with the importance of the subject, neutrino hadronization is so far largely forgotten and experiments accept bad models. This might become a serious problem in near future experiments. Although more hadronization data are coming from current and near future experiments, interpretation of these data will not be easy and the path to the future is unclear. 

\newpage

\end{document}